\documentclass[seceq]{ptptex}

\usepackage{graphicx}
\notypesetlogo                       %comment in if to eliminate PTPTeX 
\title{%        %You can use \\ for explicit line-break
The Effect of Orbital Eccentricity on Gravitational Wave Background
Radiation from Supermassive Black Hole Binaries 
}

\author{%       %Use \scshape  for the family name
Motohiro \textsc{Enoki}$^{1,}$\footnote{E-mail: enoki.motohiro@nao.ac.jp}
and Masahiro \textsc{Nagashima}$^{2,3}$%
}

\inst{%         %Affiliation, neglected when [addenda] or [errata]
$^1$National Astronomical Observatory of Japan, Mitaka
181-8588, Japan \\
$^2$Department of Physics, Kyoto University, Kyoto
606-8502, Japan \\
$^3$Faculty of Education, Nagasaki University, Nagasaki 852-8521, Japan
}

\abst{%         %this abstract is neglected when [addenda] or [errata]

A compact binary in an eccentric orbit radiates gravitational waves
(GWs) at all integer harmonics of
its orbital frequency. 
In this study, we investigate the effect of orbital eccentricity on 
the expected gravitational background radiation (GWBR) from 
supermassive black hole (SMBH) binaries in the nuclei of galaxies.
For this purpose, we formulate a power spectrum of the GWBR from
cosmological evolving eccentric
binaries. Then, we apply this formulation to the case of the GWBR from
SMBH binaries. The key to doing this 
is to correctly estimate the number density of
coalescing SMBH binaries. In this study, we use a semi-analytic
model of galaxy and SMBH formation. We find that the power spectrum of
the GWBR from SMBH binaries on eccentric orbits is
suppressed  for frequencies $f \lesssim 1~{\rm nHz}$ if the initial
eccentricity, $e_0$, satisfies $e_0 > 0.2$ 
and the initial semi-major axis is
300 times Scwarzschild radius. 
  Our model predicts that while 
the overall shape and amplitude of the 
power spectrum  depend strongly on the processes of galaxy formation,
the eccentricity of binaries can affect the shape of the power spectrum for
  lower frequencies, i.e., $f \lesssim 1~{\rm nHz}$.     
Pulsar timing measurements, which can detect GW in this
frequency range, could constrain the effect of eccentricity on the power
spectrum of the GWBR from SMBH binaries.

}

\begin{document}

\maketitle

\section{Introduction}\label{sec.intro}

An ensemble of 
gravitational waves (GWs) from a number of inspiraling 
binaries of compact objects at different redshifts can be observed
as stochastic gravitational wave background radiation (GWBR).
Coalescing supermassive black hole (SMBH) binaries with masses in the rage 
$10^6$ -- $10^9~M_{\odot}$ emit GWs, as a results of the merging of their host galaxies.
Recent studies\cite{jaffe03,enoki04} predict that these
SMBH binaries  produce GWBR of frequencies 
$\sim 1~{\rm n}$ -- $1~\mu{\rm Hz}$.
In the frequency range  $0.1~{\rm m}$ -- $10~{\rm mHz}$,
it is believed that extragalactic close binaries of white dwarfs
 are dominant sources of GWBR.\cite{farmer03}

Gravitational waves with frequencies in the range $1~{\rm n}$ -- $100~{\rm nHz}$
can be detected by pulsar timing measurements.\cite{detweiler79} \
Thus we should be able to detect the GWBR from SMBH binaries directly.
Recently, Jenet et al.\cite{jenet05} developed a new method for detecting GWBR
using multiple pulsar timing data. In the case of the Parkes Pulsar Timing Array 
(PPTA) project,\cite{hobbs05}\footnote{See http://www.atnf.csiro.au/research/pulsar/array/} 
their  results showed 
that with regular timing observations of 20 pulsars with a timing accuracy
of $100~{\rm ns}$, we should be able to directly detect the
predicted levels of the GWBR from SMBH binaries within five years. \cite{jenet05} \

In previous studies \cite{jaffe03,enoki04} of the GWBR from
SMBH binaries, it was assumed that all binaries are in circular orbits. 
However, binary orbits are generally 
eccentric. 
Before entering into the GW emitting  regime, the evolution of a SMBH binary 
is determined by its dynamical interaction with field stars in the center of its host galaxy. 
Mikkola and Valtonen \cite{mikkola92} constructed approximate
expressions relating the energy and angular momentum transfer rates for a
heavy binary and the surrounding field of light stars, and they showed
that the length of the semi-major axis decreases and the eccentricity
increases owing to dynamical friction.
Fukushige et al.\cite{fukushige92} investigated the evolution of a SMBH
binary in a uniformly distributed background of field stars and 
found that the dynamical friction on
the eccentric binary is most effective at the apocenter, where 
the orbital velocity is minimum, and thus the eccentricity increases.
Iwasawa et al.\cite{iwasawa05} performed $N$-body
simulations of the dynamical evolution of triple SMBH (a binary of SMBHs
and a single SMBH) systems in galactic
nuclei and showed that the eccentricity of the SMBH binary reaches
 $e >0.9$ in the GW emission regime. They found that both the Kozai
 mechanism and the thermalization of eccentricity due to
  the strong binary-single SMBH interaction  
drive the increase of the orbital eccentricity.  
Matsubayashi et al. \cite{matsubayashi05} investigated the orbital evolution
of intermediate mass black hole
(IMBH)-SMBH systems in galactic centers, using $N$-body
simulations. They found that the eccentricity
approaches unity ($e > 0.8$) and the IMBH can cause the SMBH to rapidly coalesce
 as a result of GW emission. Armitage and Natarajan \cite{armitage05} studied the
evolution of the SMBH binary eccentricity which is excited by the
interaction between a binary and circumbinary gas disk. 
They estimated a typical eccentricity at one week
prior to coalescence to be $e \sim 0.01$. 
In the case of extreme-mass ratio binaries, higher eccentricities ($e \sim 0.1$)
are possible.  

An eccentric binary emits GWs at all integer harmonics of
the orbital frequency.\cite{peters63,fitchett87} \  Thus, the spectral
energy distribution (SED) is different from that of a binary in a 
circular orbit, even if the masses and semi-major axes of the two binaries are
the same. 
For larger eccentricities, radiation from higher harmonics are greater and thus
the peak of the SED shifts toward higher frequency.
Moreover, the orbital evolution of a binary due to GW radiation depends strongly
on its eccentricity. As a result, the power of GW radiation of
an eccentric binary is larger than that of a circular binary, and 
the timescale of GW radiation 
 of an eccentric binary is shorter than that of a circular binary. 
Therefore, in order to predict the power spectrum of GWBR from
compact binaries, it is necessary to take account of orbital eccentricities
of binaries.  For the cases of GWBRs from galactic and extragalactic neutron star binaries
and black hole MACHO binaries, some
studies have taken account of the effect of
eccentricity.\cite{farmer03,ioka99} \ However, for the
case of the GWBR from SMBH binaries, none of models include the effect of eccentricity.      

In this study, we investigate the effect of orbital eccentricity on the 
expected GWBR from SMBH binaries.
First, we formulate the power spectrum of GWBR from cosmological eccentric binaries, taking into 
account eccentricity evolution.
In order to formulate the power spectrum, we adopt a simple
relationship between the power spectrum of the GWBR produced by 
cosmological GW sources, the total
time-integrated energy spectrum of individual sources, and the comoving
number density of GW sources found by Phinney.\cite{phinney01} 
Next, we apply this formulation to the case of the GWBR from
coalescing SMBH binaries. In order to estimate the number density of
 SMBH binaries,  we use a semi-analytic (SA) model\cite{enoki03} \  
in which SMBH formation is incorporated into galaxy formation.\cite{nagashima01}

This paper is organized as follows. 
In \S\ref{sec.binary}  
we briefly review the GW emission of a binary in an eccentric orbit. In
\S\ref{sec.gwbr} we formulate the power spectrum of GWBR from
cosmological compact binaries in eccentric orbits. In \S\ref{sec.smbh}
we present the power spectrum of the GWBR from eccentric SMBH binaries. In
\S\ref{sec.summary} we present a summary and conclusions.

\section{GWs from a binary in an eccentric orbit}\label{sec.binary}
Here we briefly review the situation for GWs emitting from a binary in an eccentric orbit 
and the evolution of a binary due to GW emission in the weak field,
slow motion limit.\cite{peters63,fitchett87}

The total power of the GW emission from a Keplerian binary consisting of
 two point masses $M_1$
 and $M_2$, with orbital frequency $f_p$ and orbital eccentricity $e$ is 
\begin{eqnarray}
L_{\rm GW}(M_1,M_2,f_p,e) & = &  L_{\rm GW, circ}(M_1,M_2,f_p) F(e), \label{eq:luminosity_e} 
\end{eqnarray}
where
\begin{eqnarray}
L_{\rm GW, circ}(M_1,M_2,f_p) &=& \frac{32}{5} \frac{G^{7/3}}{c^5}
 M_{\rm chirp}^{10/3} (2 \pi f_{p})^{10/3} \nonumber \\
 &=& 4.7 \times 10^{48} \left(\frac{M_{\rm chirp}}{10^8 \
			 M_{\odot}}\right)^{10/3}
 \left(\frac{2f_p}{10^{-7} \ {\rm Hz}} \right)^{10/3} {\rm erg},  \label{eq:luminosity_circ}
\end{eqnarray}
and
\begin{eqnarray}
F(e) & \equiv & \frac{1+73e^2/24+37e^4/96}{(1-e^2)^{7/2}}. 
\end{eqnarray}
The quantity $L_{\rm GW, circ}(M_1,M_2,f_p) $ is the total power 
from a binary in a circular orbit with masses $M_1$ and $M_2$ and 
orbital frequency $f_p$.  This is equal to the
reciprocal of the proper rest-frame period of the binary. In the above
expressions, $G$ is
the gravitational constant, $c$ is the speed of light, and 
${M_{\rm chirp}} \equiv [M_1 M_2 (M_1 + M_2)^{-1/3}]^{3/5}$ is the chirp
mass of the system. 
The distribution of the total power of GW emission 
among the harmonics of the orbital frequency is given by 
$L_{\rm GW, circ}(M_1,M_2,f_p) g(n,e)$, with the rest-frame GW frequency 
$f_r = n f_p$. 
Here, $g(n,e)$ is the GW frequency distribution function, expressed as 
\begin{eqnarray}
g(n,e) & \equiv & \frac{n^4}{32} 
\left\{\left[J_{n-2}(ne) - 2e J_{n-1}(ne)
             +\frac{2}{n}J_{n}(ne)+2eJ_{n+1}(ne)-J_{n+2}(ne) 
       \right]^2 \right. \nonumber \\
 & &  \left.    +(1-e^2)\left[
           J_{n-2}(ne) - 2e J_{n}(ne)+J_{n+2}(ne)
        \right]^2
       +\frac{4}{3n^2}\left[J_{n}(ne)\right]^2
\right\},
\end{eqnarray}
where $J_{n}$ is the $n$th-order Bessel function.
It can be shown that 
\begin{eqnarray}
\sum_{n=1}^{\infty} g(n ,e) & = & F(e).
\end{eqnarray}
For the case of a circular orbit (i.e., $e=0$), 
we have $g(2,0)=1$ and $g(n,0)=0$ for all $n \neq 2$.
The SED of gravitational radiation is given by 
\begin{equation}
L_{f_r}(e,t_p) = L_{\rm GW, circ}(f_p) \sum_{n=1}^{\infty} g(n ,e)  \delta(f_r-nf_p) \label{eq:SED_e}.
\end{equation}
Here, $t_p$ is the time at which the orbital frequency is $f_p$, and $\delta(x)$
is Dirac's delta function. 

The timescale of GW emission 
by a binary with orbital frequency $f_{p}$ measured in the rest frame is 
\begin{eqnarray}
\tau_{\rm GW} & \equiv & f_p \frac{dt_p}{df_p}. \label{eq:defgwtimescale} 
\end{eqnarray}
This timescale is given by 
\begin{eqnarray}
\tau_{\rm GW}(M_1,M_2,f_p,e) &=& \frac{\tau_{\rm GW, circ}(M_1,M_2,f_p)}{F(e)},
\end{eqnarray}
where $\tau_{\rm GW, circ}(M_1,M_2,f_p)$ is the timescale of GW emission by a binary in a circular orbit.
This timescale is given by
\begin{eqnarray}
\tau_{\rm GW, circ}(M_1,M_2,f_p)  & = & \frac{5}{96} \left(\frac{c^3}{G M_{\rm chirp}} \right)^{5/3} \left(
2 \pi f_p \right)^{-8/3}  \nonumber \\
 &=& 1.2 \times 10^4 \left(\frac{M_{\rm chirp}
			}{10^8 \ M_{\odot}}\right)^{-5/3} \left(\frac{2f_p}{10^{-7} \ {\rm Hz}} \right)^{-8/3} {\rm yr}. \label{eq:gwtimescale} 
\end{eqnarray}

\begin{figure}
\centerline{\includegraphics[width=0.42\textwidth]{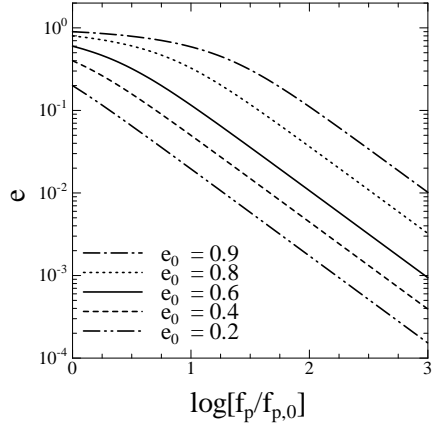}}
\caption{Evolution of the eccentricity, $e=e(f_p/f_{p,0},e_0)$, as
 a function of the orbital frequency, $f_p/f_{p,0}$, for $e_0
 = 0.2, 0.4, 0.6, 0.8$ and $0.9$. }
\label{fig01}
\end{figure}

As a result of the GW emission, the binary loses energy and angular
momentum, and as a result, the orbit of the binary becomes more circular. 
The evolutions of the semi-major axis $a$ and the eccentricity $e$ 
are described by the follows,
\begin{eqnarray}
\frac{da}{dt} &=& 
-\frac{64}{5} \frac{G^3 M_1 M_2 M_{\rm tot}}{c^5 a^3 (1-e^2)^{7/2}}
\left(1+\frac{73}{24}e^2 +\frac{37}{96}e^4 \right) \nonumber \\
 &=& -\frac{64}{5} \frac{G^3 M_1 M_2 M_{\rm tot}}{c^5 a^3} F(e),
\label{eq:a_evol} 
\end{eqnarray}
\begin{eqnarray}
\frac{de}{dt} &=& 
-\frac{304}{15} \frac{G^3 M_1 M_2 M_{\rm tot}}{c^5 a^4 (1-e^2)^{5/2}}
e \left(1+\frac{121}{304}e^2\right),\label{eq:e_evol} 
\end{eqnarray}
where $M_{\rm tot} \equiv M_1 + M_2$.
Starting from a given orbit with parameters $a_0$ and $e_0$,
Eqs. (\ref{eq:a_evol}) and (\ref{eq:e_evol}) give the following relation between $a$
and $e$:
\begin{equation}
\frac{a}{a_0} = \frac{1-e_0^2}{1-e^2} \left(\frac{e}{e_0}\right)^{\frac{12}{19}} \left[\frac{1+\frac{121}{304}e^2}{1+\frac{121}{304}e_0^2}\right]^{\frac{870}{2299}}.\label{eq:a_e_relation}
\end{equation}
From the equality $a^3=GM_{\rm tot}/(2\pi f_p)^2$, we find that the relation between the orbital
frequency and the eccentricity is given by 
\begin{equation}
\frac{f_p}{f_{p,0}} = \left\{ \frac{1-e_0^2}{1-e^2} \left(\frac{e}{e_0}\right)^{\frac{12}{19}} \left[\frac{1+\frac{121}{304}e^2}{1+\frac{121}{304}e_0^2}\right]^{\frac{870}{2299}} \right\}^{-3/2},\label{eq:f_e_relation}
\end{equation}
where $f_{p,0}$ is the initial orbital frequency.
In Fig.~\ref{fig01}, we plot the evolution of the eccentricity as a
function of the orbital frequency, $f_p/f_{p,0}$.

\section{GWBR from binaries in eccentric orbits}\label{sec.gwbr}
\subsection{Characteristic amplitude of the GWBR spectrum}

In any homogeneous and isotropic universe,
the present-day GWBR energy density, $\rho_{\rm GW} c^2$, must be equal
to the sum of
the energy densities radiated at each redshift $z$, divided by $(1+z)$ to account
for the redshifting:
\begin{eqnarray}
\rho_{\rm GW} c^2 & = & \int_0^\infty\int_0^\infty 
 n_c(z)\frac{1}{1+z}\frac{dE_{\rm GW}}{df_r} 
 df_r \, dz 
 \label{eq:edensourcea} \\
 & = & \int_0^\infty\int_0^\infty n_c(z)\frac{1}{1+z} 
        f_r\frac{dE_{\rm GW}}{df_r} \, dz \, \frac{df}{f}
 \;. 
\label{eq:edensourceb}
\end{eqnarray}
Here, $n_c(z)dz$ is the comoving number density of GW sources 
at redshifts in the range $z$ -- $z+dz$. Also $f_r$ is the GW frequency in the source's
{\it rest frame}, and $f$ is the GW frequency in the {\it observer
frame}, and thus we have $f_r=f(1+z)$. The quantity 
\begin{equation}
 \frac{dE_{\rm GW}}{df_r}df_r
\label{eq:dEdf}
\end{equation}
is the total energy emitted in GWs between the frequencies $f_r$ and $f_r+df_r$.
This energy is measured in the source's rest frame, and it is
integrated over all solid angles and over the entire radiating lifetime
of the source. 
Then, the total present-day energy density in GWs is
\begin{equation}
\rho_{\rm GW} c^2
 \equiv \int_0^\infty \frac{\pi}{4}\frac{c^2}{G} f^2 h_c^2(\ln f)\frac{df}{f}
 \;,
\label{eq:edensdef}
\end{equation}
where 
$h_c(\ln f)$ is the characteristic amplitude of the GWBR power spectrum 
over a logarithmic frequency interval $d\ln f=df/f$. 
Therefore, we find 
\begin{eqnarray}
 h_c^2(\ln f) &=& \frac{4 G}{\pi c^2 f^2} \int_0^\infty n_c(z)\frac{1}{1+z} 
 \left.\left(f_r\frac{dE_{\rm GW}}{df_r}\right)\right|_{f_r=f(1+z)} dz, \nonumber \\
&=& \frac{4 G}{\pi c^2 f} \int_0^\infty n_c(z)
 \left.\left(\frac{dE_{\rm GW}}{df_r}\right)\right|_{f_r=f(1+z)} dz. \label{eq:spectrum}
\end{eqnarray} 
If the GW sources are coalescing binaries, the characteristic amplitude
is given by
\begin{eqnarray}
 h_c^2(\ln f)
&=& \frac{4 G}{\pi c^2 f} \int dM_1 dM_2 dz \; n_c(M_1,M_2,z)
 \left.\left(\frac{dE_{\rm GW}(M_1,M_2)}{df_r}\right)\right|_{f_r=f(1+z)}. 
\label{eq:chracter_s},
\end{eqnarray} 
where $n_c(M_1,M_2,z)dM_1 dM_2 dz$ is the comoving number density of
 coalescing binaries with masses $M_1$ -- $M_1 + dM_1$ and 
$M_2$ -- $M_2 +dM_2$ for $z$ -- $z +dz$.

Note that Eq. (\ref{eq:spectrum}) assumes that the timescale of GW
emission is much shorter  than the Hubble time.  
This relationship between the power spectrum of the GWBR produced by 
cosmological GW sources and the SED of an individual source was
 derived by Phinney.\cite{phinney01}

\subsection{Total energy emitted in GWs}

The total energy of GW radiation emitted over a lifetime of duration $t_{\rm life}$ is  
\begin{eqnarray}
E_{\rm GW} &=& \int_{0}^{t_{\rm life}} L_{\rm GW}(t_p) dt_p \\
&=& \int_{0}^{t_{\rm life}} \int L_{f_r}(t_p) df_r dt_p,
\end{eqnarray}
where $L_{\rm GW}(t_p)$ is the power of GW emission and $L_{f_r}(t_p)$
is the SED of GW emission. Then, we have
\begin{equation}
\frac{dE_{\rm GW}}{df_r} =  \int_{0}^{t_{\rm life}} L_{f_r}(t_p) dt_p.\label{eq:gwSED}
\end{equation}
Therefore, given the number density and SED of GW sources, we can
calculate the amplitude of the GWBR power spectrum. 

\subsection{Characteristic amplitude of the GWBR spectrum from eccentric binaries}

\begin{figure}
\centering\includegraphics[scale=0.42]{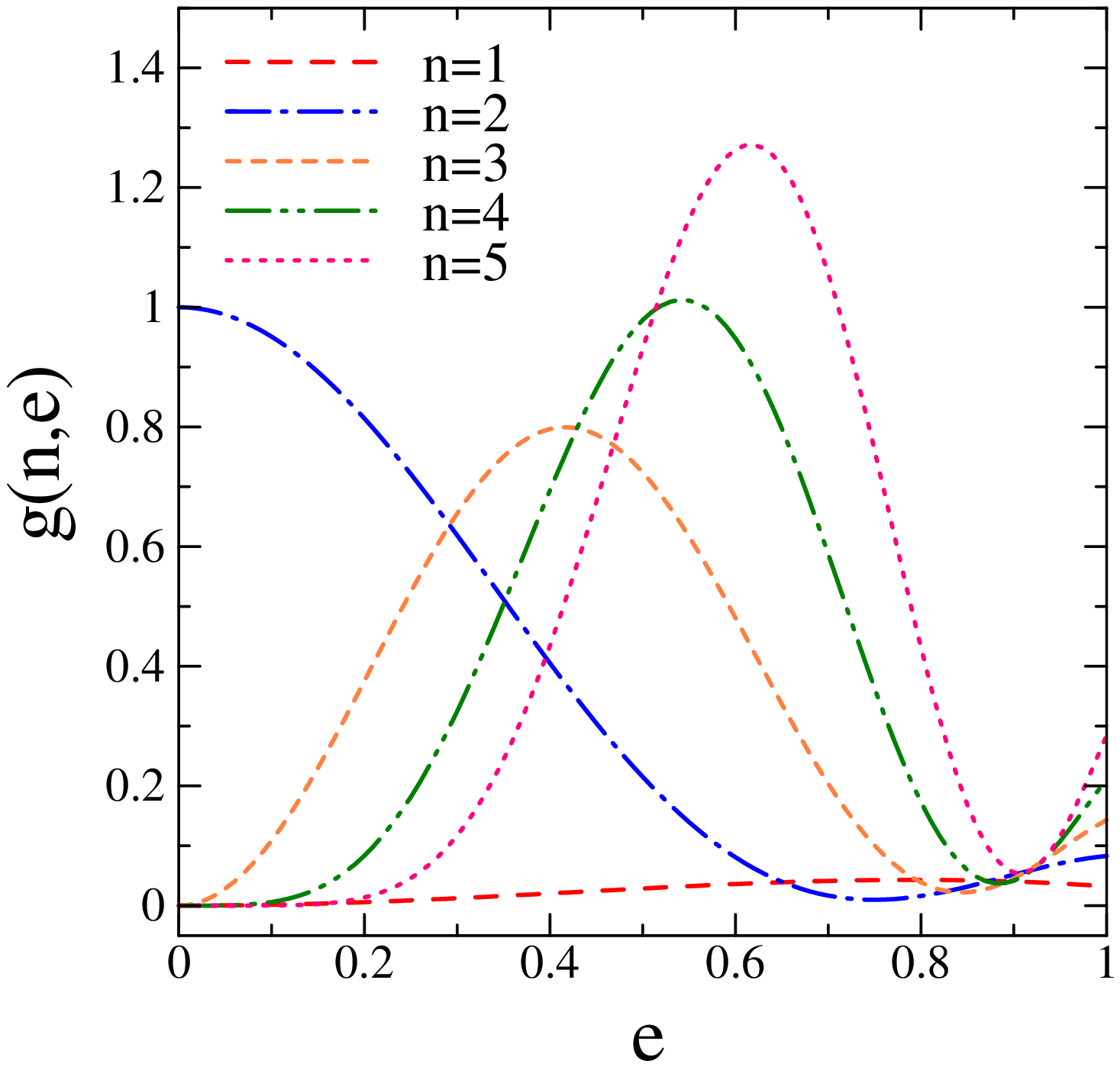}
\centering\includegraphics[scale=0.42]{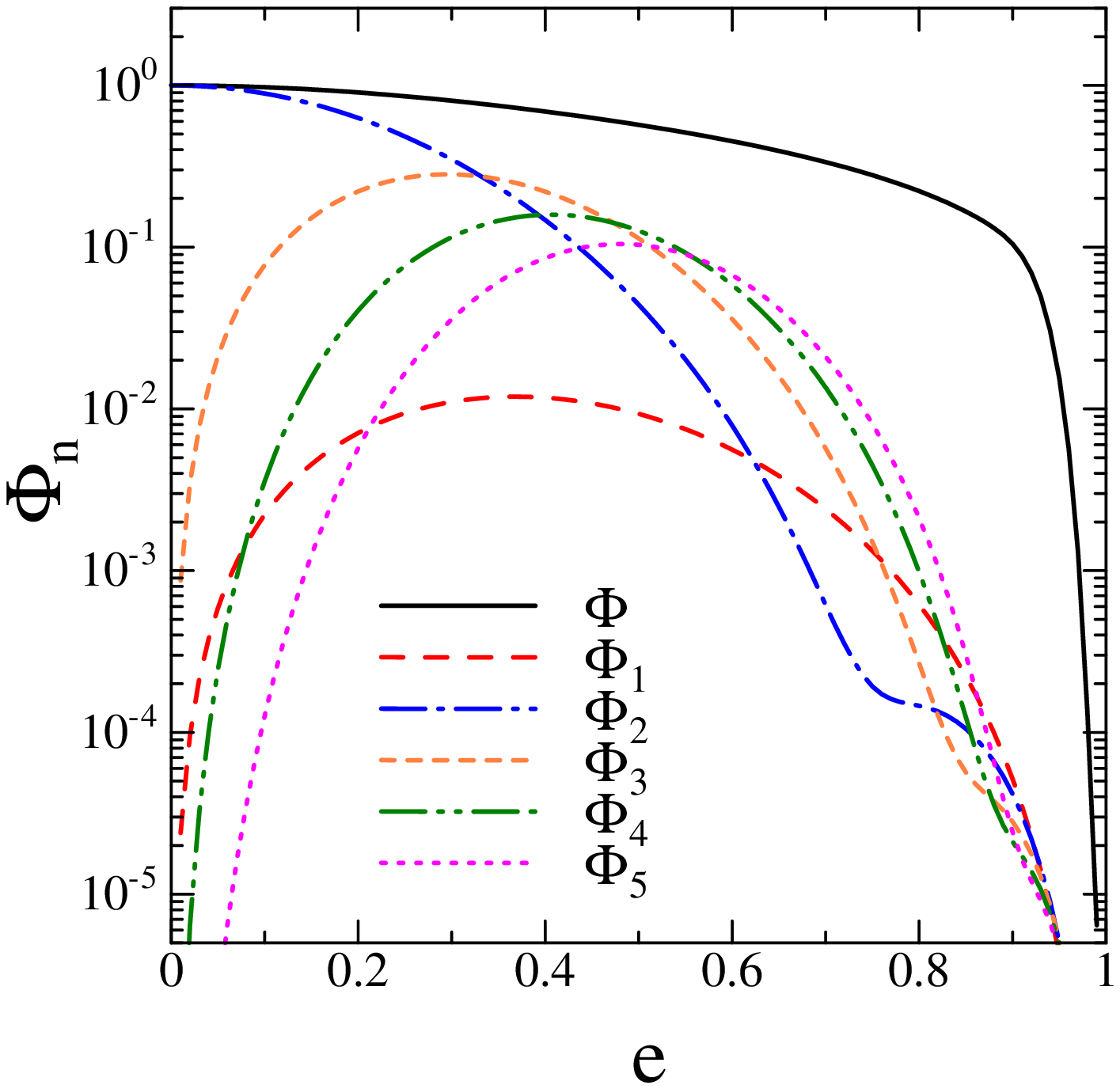}
\caption{Left panel: $g(n,e)$ as
 a function of $e$ for $n=1,2,3,4$ and $5$.  Right panel: $\Phi$ and
 $\Phi_n$ for $n=1,2,3,4$ and $5$ as
 functions of $e$.}
\label{fig02}
\end{figure}

For a Keplerian binary consisting of two point masses $M_1$
 and $M_2$ with an orbital eccentricity $e$, the
 SED of the GW radiation is given by Eq. (\ref{eq:SED_e}).
From Eqs. (\ref{eq:SED_e}), (\ref{eq:defgwtimescale}) and (\ref{eq:gwSED}), we obtain
\begin{eqnarray}
\frac{dE_{\rm GW}}{df_r} &=& \int L_{\rm GW, circ}(f_p) \frac{dt_p}{df_p} \sum_{n=1}^{\infty} g(n ,e)  \delta(f_r-nf_p)  df_p \nonumber \\
&=& \sum_{n=1}^{\infty} \left.\left[L_{\rm GW, circ}(f_p)
			       \frac{\tau_{\rm GW}(f_p,e) }{n f_p} g(n,e) \right]\right|_{f_p = f_r/n} . \label{eq:ISED_e}
\end{eqnarray}
Thus, from Eqs. (\ref{eq:chracter_s}) and (\ref{eq:ISED_e}), the
power spectrum of
GWBR from  binaries in eccentric orbits is given by 
\begin{eqnarray}
 h_c^2(\ln f)
&=& \frac{4 G}{\pi c^2 f} \int dM_1 dM_2 dz \; n_c(M_1,M_2,z)
\sum_{n=1}^{\infty} \left.\left[L_{\rm GW, circ}(f_p) \frac{\tau_{\rm
			   GW}(f_p,e) }{n f_p} g(n,e)
			  \right]\right|_{f_p = f(1+z)/n} \nonumber\\
 &=&\frac{4 \pi c^3}{3} \int dM_1 dM_2 dz \; n_c(M_1,M_2,z) (1+z)^{-1/3}  
\left(\frac{GM_{\rm chirp}}{c^3}\right)^{5/3}
 (\pi f)^{-4/3}  \Phi \label{eq:strain_e}, 
\end{eqnarray} 
where 
\begin{eqnarray}
\Phi & \equiv &  \sum_{n=1}^{\infty} \Phi_n 
\end{eqnarray}
and
\begin{eqnarray}
\Phi_n & \equiv &  \left(\frac{2}{n}\right)^{2/3} \frac{g(n,e)}{F(e)}.
\end{eqnarray}
The left panel of Fig. \ref{fig02} plots $g(n,e)$ as a function of
the eccentricity, $e$, and the right panel plots $\Phi$ and $\Phi_n$  
as functions of $e$.

\begin{figure}
\centering\includegraphics[scale=0.42]{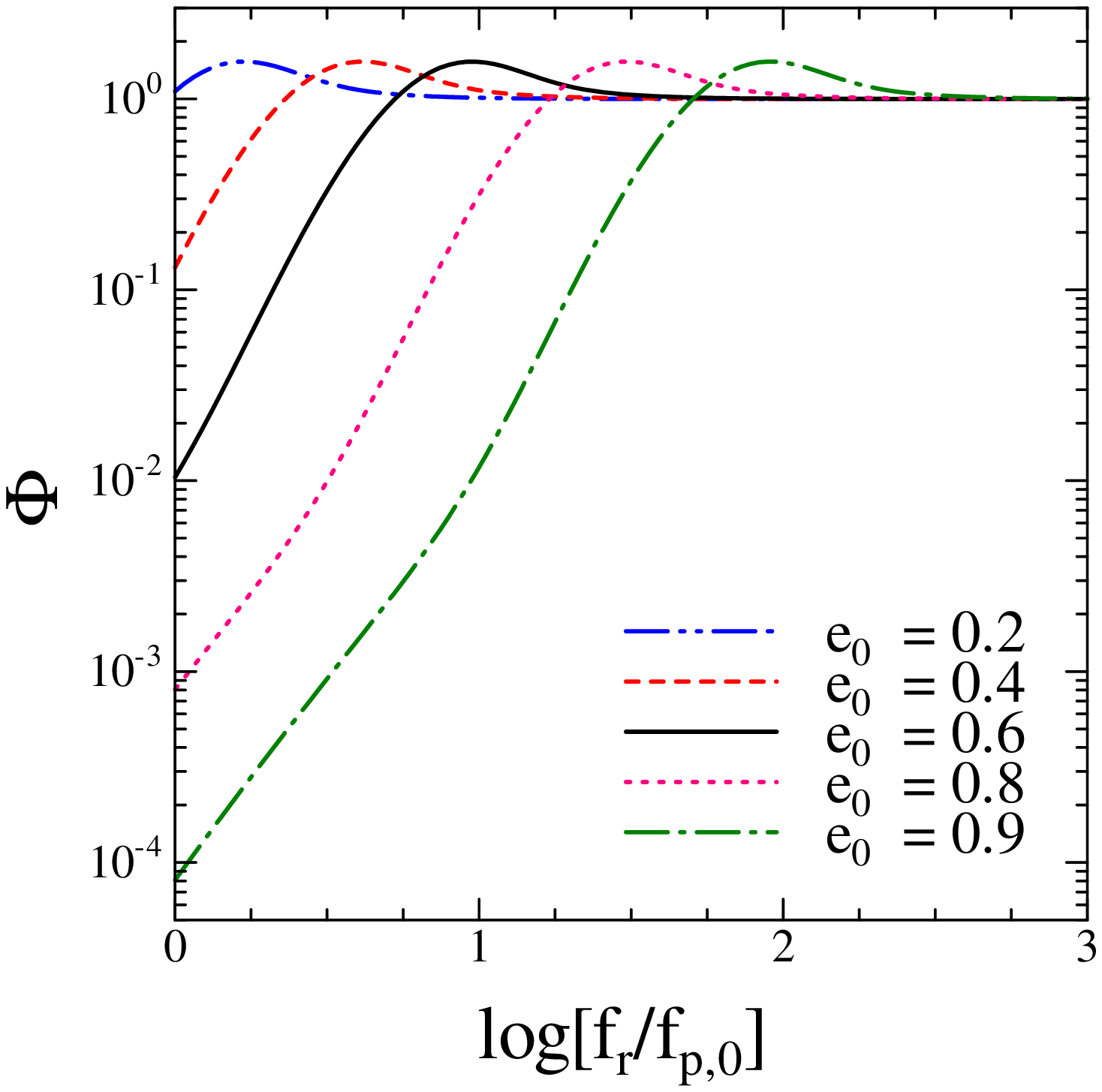}
\centering\includegraphics[scale=0.42]{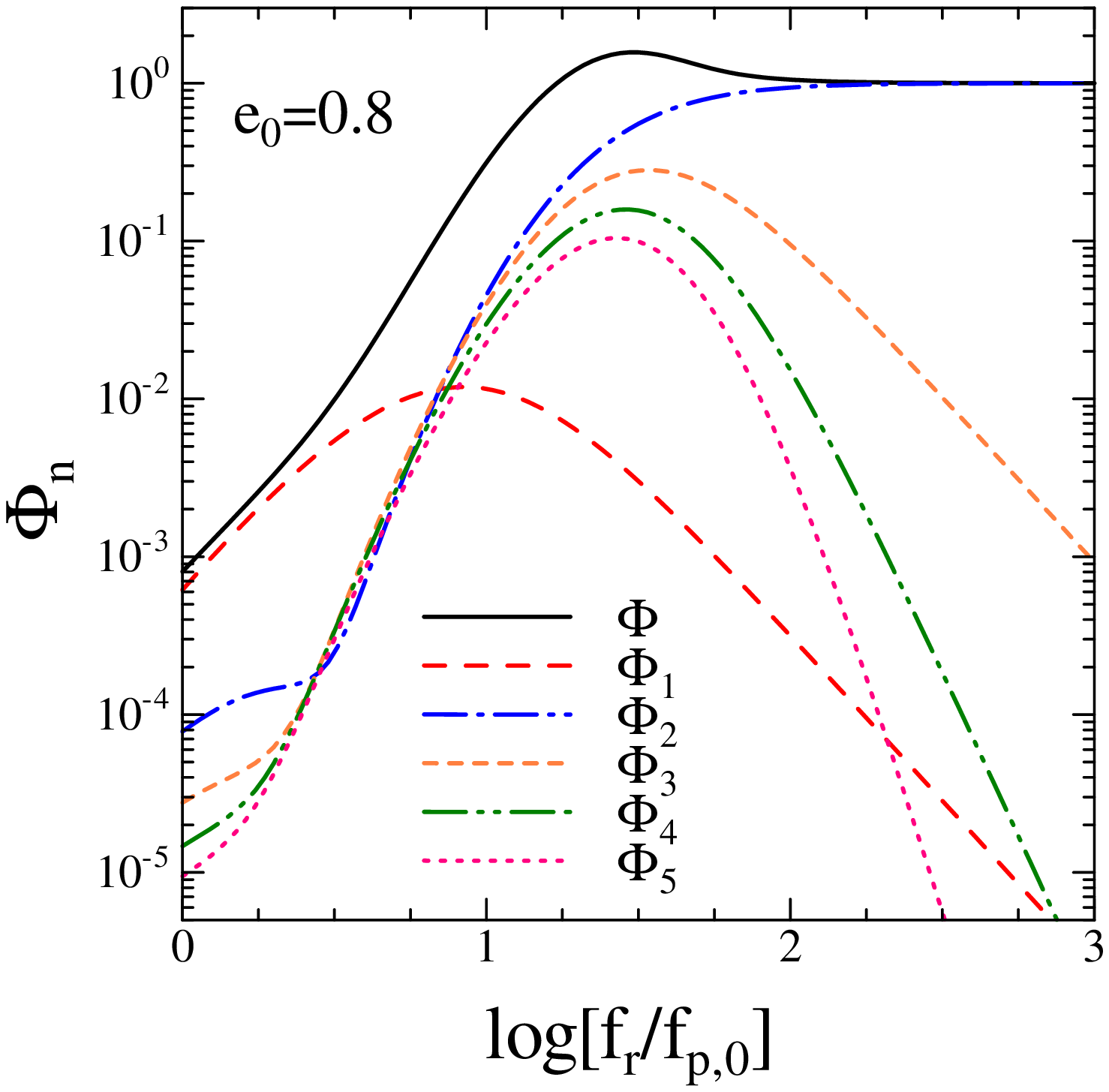}
\caption{ Strength
 of the effect of the orbital
 eccentricity on the GWBR power spectra.
Left panel: $\Phi =\Phi(f_r/f_{p,0},e_0)$ as
 a function of $f_r/f_{p,0}$ for 
$e_0 = 0.2, 0.4, 0.6, 0.8$ and $0.9$.  Right panel: $\Phi_n$ with $e_0=0.8$ as
 a function of $f_r/f_{p,0}$ for $n=1,2,3,4$ and
 $5$. $\Phi$ for  $e_0=0.8$ is also shown.}
\label{fig03}
\end{figure}

Here we note that the eccentricity is a function of the orbital
frequency, $f_p/f_{p,0}$, [see
Eq. (\ref{eq:f_e_relation}) and Fig. \ref{fig01}]. 
Therefore, $e=e(f_p/f_{p,0},e_0)=e(f_r/nf_{p,0},e_0)=e[f(1+z)/nf_{p,0},e_0]$.
Hence, $\Phi$ depends on $f$, $f_{p,0}$, $e_0$ and $z$.
 Since SMBH binaries have various initial
eccentricities, the number density in Eq. (\ref{eq:strain_e}) should
be $n_c(M_1,M_2,e_0,z)$ and it should be integrated over $e_0$. 
If all binaries are circular (i.e., $e_0=0$), $\Phi_2 = 1$ and $\Phi_n = 0$
 for $n \ne 2$. Thus, we have $ h_c \propto f^{-2/3}$ for $e_0=0$.
 By contrast,
 for $e_0 \ne 0$, $\Phi$ depends on $f$, and $h_c$ is not
 proportional to $f^{-2/3}$.  Therefore, $\Phi$ indicates the strength
 of the effect of orbital
 eccentricity on the GWBR power spectrum.
In the left panel of Fig. \ref{fig03}, we plot $\Phi$  as
 a function of rest-frame GW frequency, $f_r/f_{p,0}$, for some initial
 eccentricities. 
In the right panel of Fig. \ref{fig03}, we plot
 $\Phi_n$ with $e_0=0.8$ as a function of $f_r/f_{p,0}$ for some harmonics.
Owing to the radiation of the harmonics of $f_{p}$, the power spectrum
is suppressed
 at lower frequencies, i.e. $f_r/f_{p,0} \lesssim 10$, and it is amplified at intermediate
frequencies, i.e.  
$10 \lesssim f_r/f_{p,0} \lesssim 100$.
For larger frequencies, $100 \lesssim f_r/f_{p,0}$, the main contributors
to the power spectrum are circular binaries, which evolve from eccentric
binaries and radiate only through the $n =2$ mode.

\section{Power spectrum of the GWBR from eccentric SMBH binaries}\label{sec.smbh}
\subsection{A unified semi-analytic model of galaxy and SMBH formation}

In order to predict the power spectrum of the GWBR from coalescing SMBH binaries,
we must estimate the number density of coalescing SMBH binaries.  To do this, we
use a semi-analytic (SA) model\cite{enoki03} \  
in which SMBH formation is incorporated into galaxy formation.\cite{nagashima01}

In the standard hierarchical structure formation scenario in a cold dark
 matter (CDM) universe, dark-matter halos ({\it dark halos}) cluster
 through the effect of gravity
and merge. In each merged dark halo, a galaxy 
is formed as a result of  
radiative gas cooling. In each galaxy, star formation and  
supernova feedback occur. 
Several galaxies in a common dark halo sometimes merge, and a
more massive galaxy is thereby formed. When galaxies merge, SMBHs in 
the centers of these galaxies move toward the center of the new merged galaxy and 
 subsequently form a SMBH binary.
If the binary loses a sufficient amount of  energy and angular 
momentum, it will evolve into the GW emitting 
regime and begin inspiraling, eventually coalescing with a 
GW burst. 

In SA models, merging histories of dark  halos are realized  using a
Monte-Carlo algorithm, and the evolution of baryonic components within
dark halos is calculated using simple analytic models for gas cooling,
star formation, supernova feedback, galaxy merging and other processes. 
SA models have successfully reproduced a variety of observed
 features of galaxies, such as their luminosity functions and gas
 fractions in disk galaxies. In our SA model in which SMBH formation has
 been incorporated,\cite{enoki03} \  it is assumed that 
SMBHs grow through the coalescence of two or more SMBHs when
 their host galaxies merge.
We also assume that during a major merger, a fraction of the cold gas 
proportional to the total mass of stars newly formed at the starburst is
accreted onto the newly formed SMBH and that this gas fueling leads to quasar
activity. Thus, SMBHs also grow through the accretion of cold gas.
Our SA model reproduces not only observational features of galaxies
\cite{nagashima01} but also  
the present-day observed SMBH mass function and the quasar luminosity functions at different
redshifts.\cite{enoki03} Using this SA model, Enoki et al. estimated
the amplitude of GWBR from inspiraling SMBH binaries in circular
orbits and the event rate of GW bursts due to SMBH binary coalescing.\cite{enoki04}   

It is difficult to determine how SMBH binaries manage to shrink into a
regime in which their evolution is driven by 
GW emission  after their host galaxies merge, because  not all
the physical processes and conditions related to this problem (dynamical
friction, the stellar distribution, triplet SMBH interaction, gas dynamical
effects, and so on) are yet clear
(see, e.g. Refs. \cite{mikkola92,fukushige92} and \cite{begelman80}). 
Therefore, it is also difficult to predict
 the distribution of the initial eccentricity. 
In what follows, for simplicity, we assume that all SMBH binaries coalesce when host
galaxies merge. In other words, the efficiency 
of SMBH coalescence is assumed to be maximal.
 Thus, the efficiency of SMBH coalescence is maximal
and the predicted amplitude of the GWBR spectrum should be
interpreted as the upper limit (see Enoki et al.\cite{enoki04}).   
 In this paper, in order to clarify the effect of eccentricity, we 
 assume 
that all binaries have a
 single initial eccentricity given at a radius that is equal to some
 constant times the Schwarzschild radius of the SMBH.

In this study, the adopted cosmological model is a low-density,
 spatially flat cold dark matter ($\Lambda$CDM) universe with
the present density parameter, $\Omega_{\rm m}=0.3$, the cosmological
constant $\Omega_{\Lambda}=0.7$, and the Hubble constant $h=0.7$ 
($h \equiv H_0/100 \; {\rm km \ s^{-1}\ {Mpc^{-1}}}$). A detailed 
 description of the model and its parameters is given in Nagashima et
 al.\cite{nagashima01} and  Enoki et al.\cite{enoki03,enoki04}

\subsection{Maximum orbital frequency}

Substituting  the number density of coalescing SMBH binaries,
$n_c(M_1,M_2,z)dM_1 dM_2 dz$, into Eq. (\ref{eq:strain_e}),
 we can calculate the power spectrum of the GWBR from SMBH coalescing SMBH
 binaries.
Here, we introduce the cut off orbital frequency
$f_{p, {\rm max}}$.\cite{enoki04} \  As a  binary evolves with time
owing to its emission of GW, the 
frequency increases. We assume that 
the binary orbit is quasi-stationary until
the radius equals $3 R_{\rm S}$, where $R_{\rm S}$ is the Schwarzschild
radius, i.e. the radius of the innermost stable circular orbit (ISCO) for
a particle around a non-rotating black hole. 
Then, the maximum orbital frequency $f_{p, {\rm max}}$ is 
\begin{eqnarray}
f_{p, {\rm max}}(M_1,M_2) &=& \frac{c^3}{6^{3/2} ~ 2 \pi G M_1} \left(1+\frac{M_2}{M_1} \right)^{1/2} \nonumber \\
&=& 2.2 \times 10^{-5} \left(\frac{M_1}{10^8 M_{\odot}}\right)^{-1} \left(
1+\frac{M_2}{M_1} \right)^{1/2} {\rm Hz}, \label{eq:fmax}
\end{eqnarray}
where $M_1$ and $M_2$ are the SMBH masses (and we assume $M_1 > M_2$).
Then, in Eq. (\ref{eq:SED_e}), we replace $L_{\rm GW, circ}(f_p)$
by $L_{\rm GW, circ}(f_p) \theta (f_{p, {\rm max}} - f_p)$, where
$\theta(x)$
 is the step function. Therefore, $\Phi_n$ becomes
\begin{eqnarray}
\Phi_n & = &  \left(\frac{2}{n}\right)^{2/3} \frac{g(n,e)}{F(e)} \theta
 \left( f_{p, {\rm max}} - f(1+z)/n \right).
\end{eqnarray}
If this cut off does not exist, we have $h_c \propto f^{-2/3}$ for $f \gtrsim 2 f_{p, {\rm max}}$. 

\subsection{Results}

\begin{figure}
\centering\includegraphics[scale=0.42]{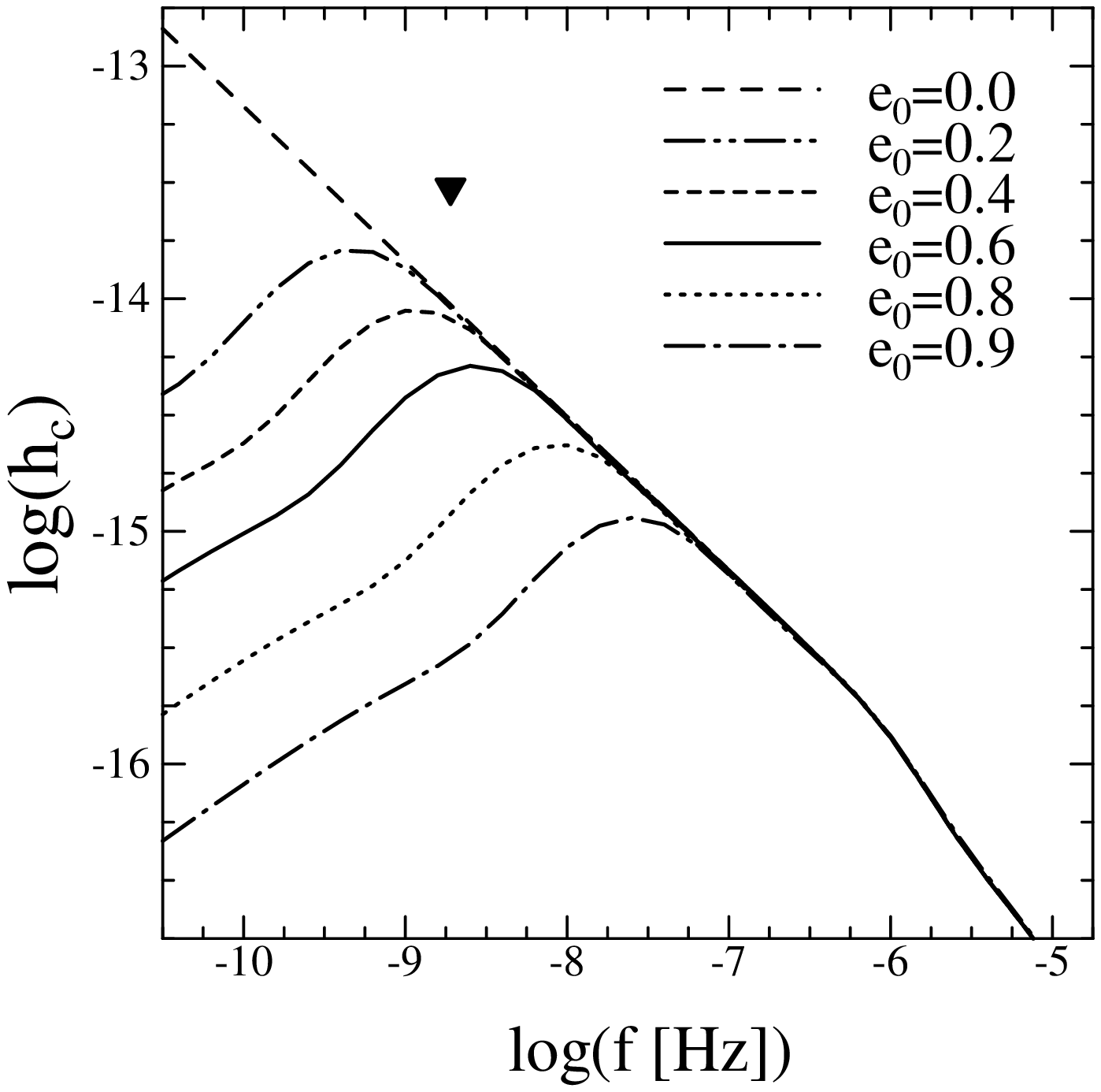}
\centering\includegraphics[scale=0.42]{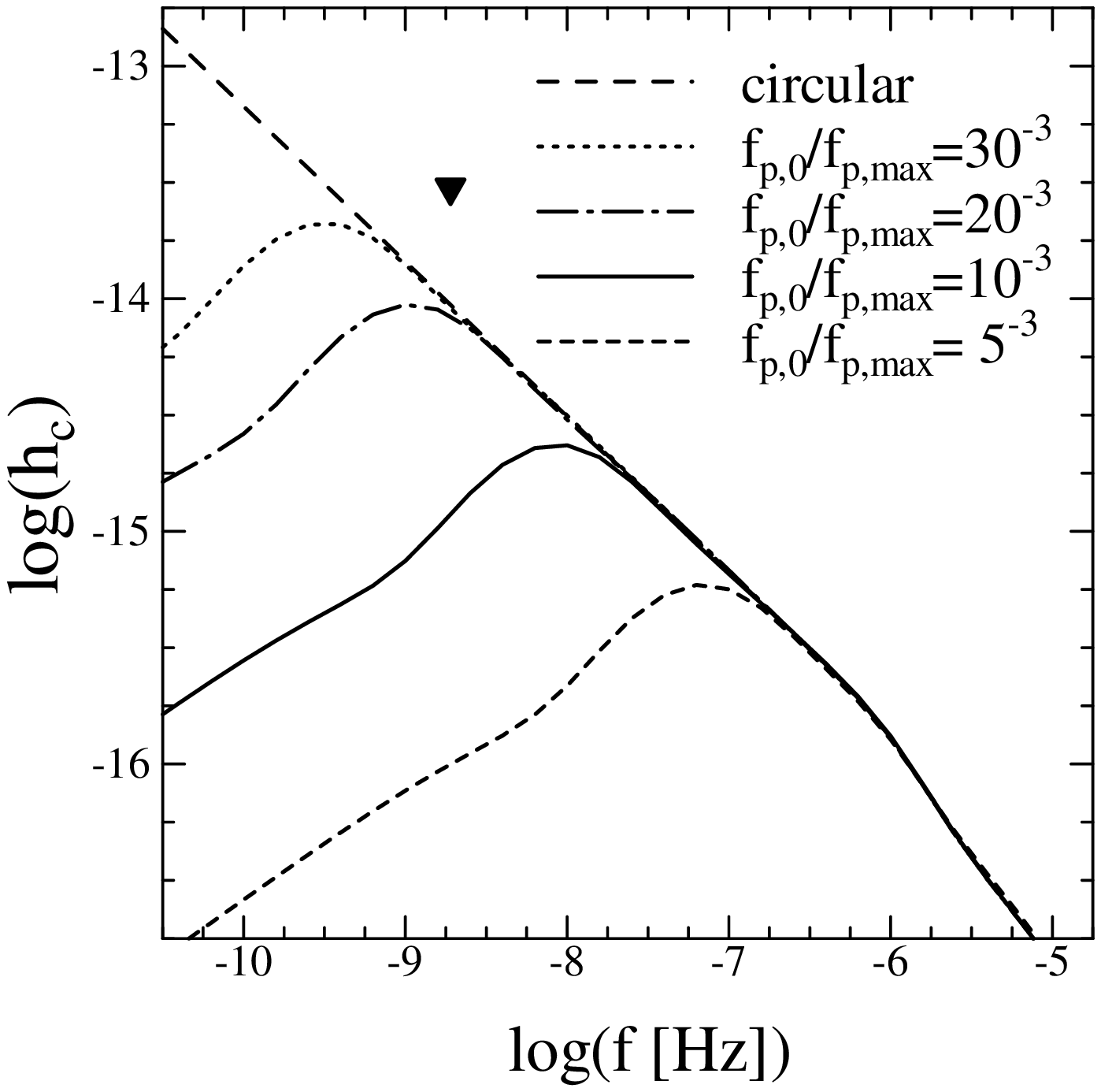}
\caption{Characteristic amplitudes of GWBR power spectra over a
 logarithmic frequency interval,
 $h_{c}(\ln f)$, from SMBH binaries. Left panel: 
Power spectra of
 GWBR from SMBH binaries with $f_{p,0}/f_{p, {\rm max}} = 10^{-3}$ for
several initial eccentricities, $e_0 = 0.0, 0.2, 0.4, 0.6, 0.8$ and $0.9$. 
Right panel: Power spectra of
 GWBR from SMBH binaries for $e_0=0.8$ for several initial orbital
 frequencies, $f_{p,0}/f_{p, {\rm max}} = 5^{-3}, 10^{-3}$, $20^{-3}$ and $30^{-3}$.
In each panel, the solid triangle
 indicates the current limit from pulsar timing measurements.\cite{lommen02}}
\label{fig04}
\end{figure}

In the left panel of Fig. \ref{fig04}, we plot power spectra of GWBR, $h_{c}$,
from SMBH binaries for various initial eccentricities.  The initial
eccentricities, $e_0$, are those at $f_{p,0}/f_{p, {\rm max}} = 10^{-3}$ ($a = 300R_{\rm S}$).  
This figure shows that power spectra at the frequencies measured by pulsar
timing ($\sim 1~{\rm n} - 100~{\rm nHz}$) are suppressed as a result of harmonic
radiation, especially for $e_{0} \gtrsim 0.4$.
The slope of the spectrum changes near $f = 1~\mu{\rm Hz}$, owing to a lack of
power associated with the upper limit frequency, $f_{p, {\rm max}}$.

The right panel of Fig. \ref{fig04} displays power spectra of
 GWBR from SMBH binaries for $e_0=0.8$ for different  initial orbital
 frequencies with $f_{p,0}/f_{p, {\rm max}} = 5^{-3}, 10^{-3}$, $20^{-3}$ and $30^{-3}$.
 Because $a \propto f_{p}^{-2/3}$, these correspond to $a = 75~R_{\rm
 S}, 300~R_{\rm S}, 1200~R_{\rm S}$
 and $2700~R_{\rm S}$, respectively. 
Even in the case $f_{p,0}/f_{p, {\rm max}} = 30^{-3}$, the power
 spectrum for $f \lesssim 1~{\rm nHz}$ is suppressed.
We note that for a binary with $e_0=0.8$ at $f_{p,0}/f_{p, {\rm max}} = 30^{-3}$
($a = 2700 R_{\rm S}$), the eccentricity of the binary is $e \sim 0.95$
 for $a \sim 10^4~R_{\rm S} \sim 0.1~(M_{\rm BH}/10^8 M_{\odot})~{\rm pc}$. 
Both panels of Fig. \ref{fig04} indicate that the shape
 of the power
 spectrum in the low frequency range depends strongly on $e_0$ and $f_{p,0}/f_{p, {\rm max}}$. 

\begin{figure}
\centerline{\includegraphics[width=0.42\textwidth]{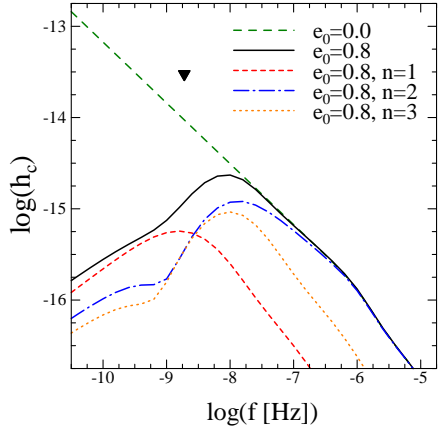}}
\caption{Power spectra of GWBR from SMBH binaries for $e_0=0.8$. 
Some harmonics of the GWBR spectrum for $n=1,2$ and $3$. The initial orbital
 frequency is  $f_{p,0}/f_{p, {\rm max}} = 10^{-3}$.
We also plot the power spectrum of GWBR from circular binaries (dashed line)
 and the current limit from pulsar timing measurements
 (solid triangle).\cite{lommen02} }
\label{fig05}
\end{figure}

In Fig. \ref{fig05}, we plot some harmonics of the GWBR
power spectrum from SMBH binaries for  $e_0=0.8$ and  $f_{p,0}/f_{p, {\rm max}} = 10^{-3}$.
As with the right panel of Fig. \ref{fig03}, in the lower frequency range,
$f \lesssim 1~{\rm nHz}$, the $n=1$ mode is dominant
 and in the higher frequency range, $f \gtrsim 0.1~\mu{\rm Hz}$, the $n=2$ mode is dominant. 
Therefore, in this frequency range, GWs from circular binaries are dominant.

\begin{figure}
\centerline{\includegraphics[width=0.45\textwidth]{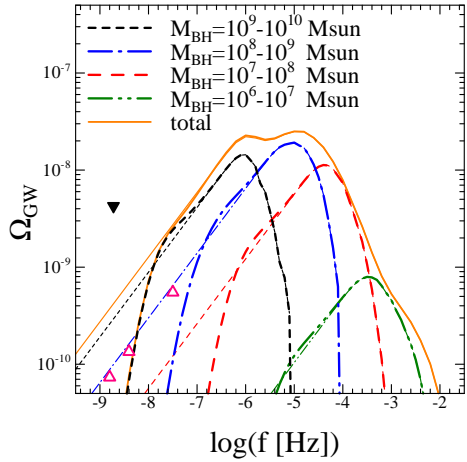}}
\caption{Energy density parameters of GWBR over a logarithmic frequency
 interval $\Omega_{\rm GW}(\ln f)$ in several intervals of the  total
 mass. The thick curves represent the results for $e_0=0.8$ and
 $f_{p,0}/f_{p, {\rm max}} = 10^{-3}$, and thin curves represent the
 results for $e_0 = 0$. The solid triangle
 indicates the current limit from pulsar timing
 measurements.\cite{lommen02}\
 The three other triangles indicate the potential future lower limits from the
 full PPTA data set for the case $h_c \propto f^{-2/3}$.\cite{jenet06} }
\label{fig06}
\end{figure}

The present-day energy density parameter of GWBR over a logarithmic frequency interval
$d\ln f=df/f$ is given by 
\begin{eqnarray}
\Omega_{\rm GW}(\ln f) &=& \frac{2 \pi^2}{3 H_0^2} f^2 h_c^2(\ln f).
\end{eqnarray}
In Fig. \ref{fig06}, we plot 
$\Omega_{\rm GW}(\ln f)$ from SMBH binaries with $e_0=0.8$
and  $f_{p,0}/f_{p, {\rm max}} = 10^{-3}$ and with $e_0=0$ for several
intervals of the  total SMBHs mass ($M_{\rm tot} = M_1+ M_2$).
It is seen that the main contribution to the energy density is from binaries with total masses
$M_{\rm tot} = 10^7 - 10^{10} M_{\odot}$. In the lower frequency range
($f \lesssim 0.1~\mu {\rm Hz}$), which
can be measured by pulsar timing, GWs from binaries with masses $M_{\rm tot}
\ge 10^9 M_{\odot}$ are dominant.
As with the left panel of Fig. \ref{fig03}, Fig. \ref{fig06} shows 
suppressions and amplifications of the energy density parameters
for each SMBH mass interval due to the harmonic radiation.
Here, the initial eccentricity is that at $f_{p,0}/f_{p, {\rm max}} = 10^{-3}$. 
From Eq. (\ref{eq:fmax}), we find 
$f_{p,0} \propto f_{p, {\rm max}} \propto (M_{1} M_{\rm tot})^{-1/2}$. 
Thus, for smaller-mass SMBHs,
non-zero eccentricities affect the power of higher frequencies. 
When we sum up the energy densities of GWBR from SMBH mass intervals at approximately
$10~{\rm nHz}$, where the energy density of GWBR from SMBH binaries with
 $M_{\rm tot} > 10^9~M_{\odot}$ increases,
 the energy density of GWBR from SMBH binaries with  $M_{\rm tot} < 10^9~M_{\odot}$ 
has decreased significantly.   
For this reason, we cannot observe the amplification of the total energy density
of GWBR from SMBH binaries.

Pulsar timing measurements allow GWs with frequencies in the range 
$1~{\rm n} - 100~{\rm nHz}$ to be detected.
Under the assumption that $h_c \propto f^{-2/3}$,
the completed PPTA data set (twenty pulsars with an rms timing residual of
100 ns over 5 years) could potentially provide limits on  the level of GWBR from
SMBH binaries: $\Omega_{\rm GW} =  5.5 \times 10^{-10}$ at 
$f = 10^{-7.5}~{\rm Hz} ~(1 ~{\rm yr}^{-1}) $,
$\Omega_{\rm GW} = 1.3 \times 10^{-10}$ at $f = 10^{-8.4}~{\rm Hz} ~(1/8 ~{\rm yr}^{-1})$
and 
$\Omega_{\rm GW} = 7.3 \times 10^{-11}$ at $f = 10^{-8.8}~{\rm Hz}~(1/20 ~{\rm yr}^{-1})$.\cite{jenet06} \ 
In Fig. \ref{fig06}, we also plot the potential future limits. 
These limits are well under the predicted energy density of GWBR from SMBH
binaries for $e_0 = 0$ ($h_c \propto f^{-2/3}$).
Therefore, we conclude that the full PPTA data set should be sufficient to constrain
the effect of eccentricity on the GWBR from SMBH binaries.

\subsection{Effects of galaxy formation processes on GWBR from SMBH binaries}

The number density of coalescing SMBH
binaries with masses in the range $M_1 - M_1 + dM_1$ and 
$M_2 - M_2 +dM_2$ at $z - z +dz$, i.e.
$n_c(M_1,M_2,z)dM_1 dM_2 dz$, depends on galaxy formation
processes. In our SA model, the dominant mass growth process of SMBHs is
the accretion of cold gas, which is also the material of
stars\cite{enoki04}. Thus, $n_c(M_1,M_2,z)$ depends strongly on
processes related to star formation related.
 Here, we show how the star formation time scale
and the strength of supernova feedback affect the power spectrum of the 
GWBR from SMBH binaries.

In our SA model, the star formation rate (SFR) of a galaxy
is assumed to be $\dot{M}_{*}={M_{\rm cold}}/{\tau_{*}}$, 
where $M_{\rm cold}$ is the mass of the cold gas and
$\tau_{*}$ is the time scale of star formation. We assume 
$\tau_{*}=\tau_{*}^{0} (V_{\rm circ}/300~{\rm km~s}^{-1})^{\alpha_{*}}$,
 where $V_{\rm circ}$ is the circular velocity of the galaxy. 
The free parameters $\tau_{*}^{0}$ and $\alpha_{*}$ 
are chosen to match the observed mass fraction of cold gas in the disks
of spiral galaxies. 
With star formation, supernovae occur
and heat up the surrounding cold gas, yielding a hot gas phase (a 
process called supernova feedback). The reheating rate is given by 
 $\dot{M}_{\rm reheat}=\beta(V_{\rm circ}) \dot{M}_{*}$, where
 $\beta(V_{\rm circ}) = (V_{\rm hot}/V_{\rm circ})^{\alpha_{\rm hot}}$. 
The free parameters $V_{\rm hot}$ and $\alpha_{\rm hot}$ 
are determined by matching the observed local luminosity function of
galaxies. In a previous study\cite{enoki04} (and in a previous subsection of
this paper), we adopted
 $\tau_{*}^{0} = 1.5~{\rm Gyr}, \alpha_{*} = -2, V_{\rm hot} = 280~{\rm
 km~s}^{-1}$  and $\alpha_{\rm hot} = 2.5$ as fiducial values. 
 
\begin{figure}
\centering\includegraphics[scale=0.42]{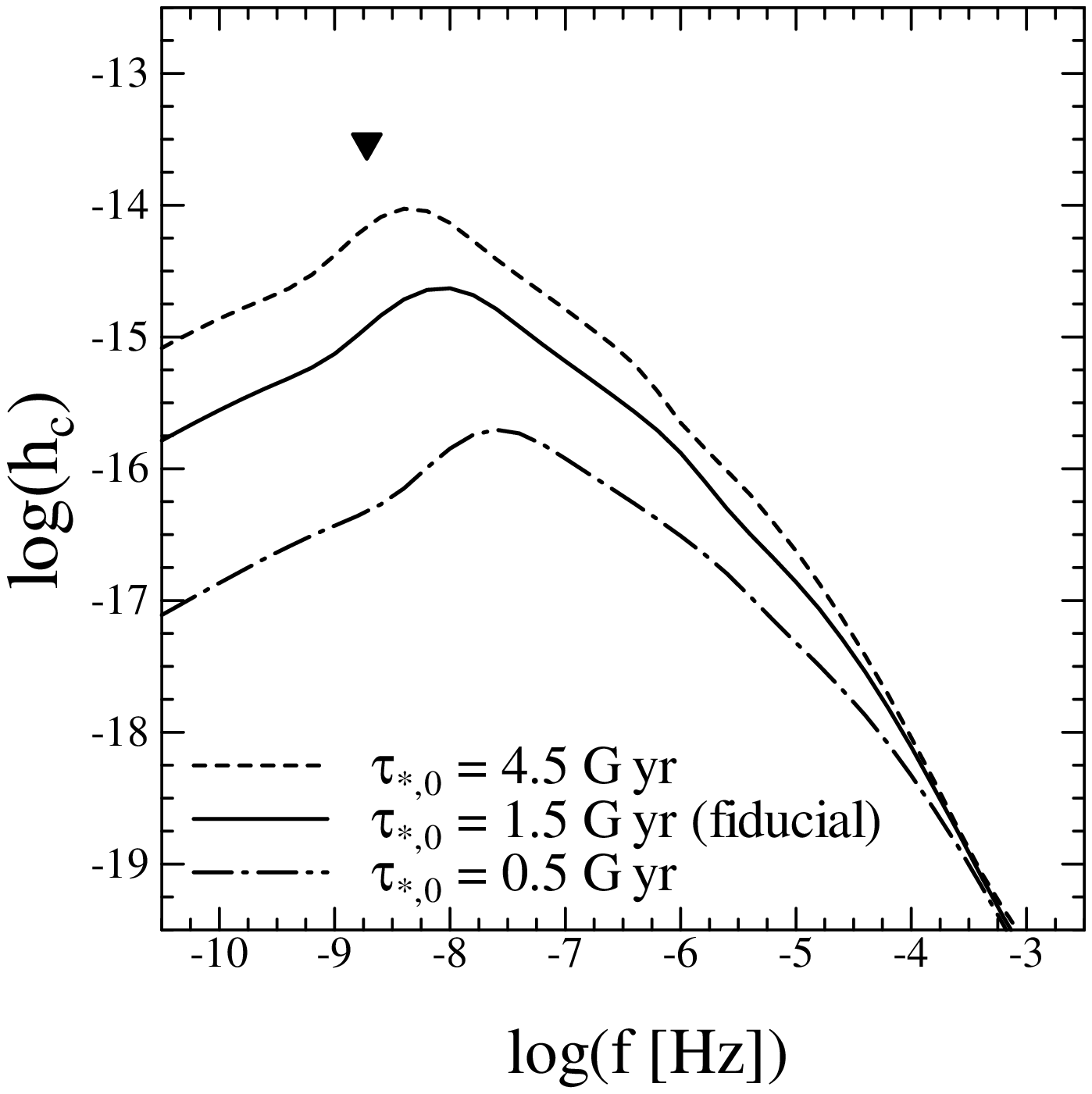}
\centering\includegraphics[scale=0.42]{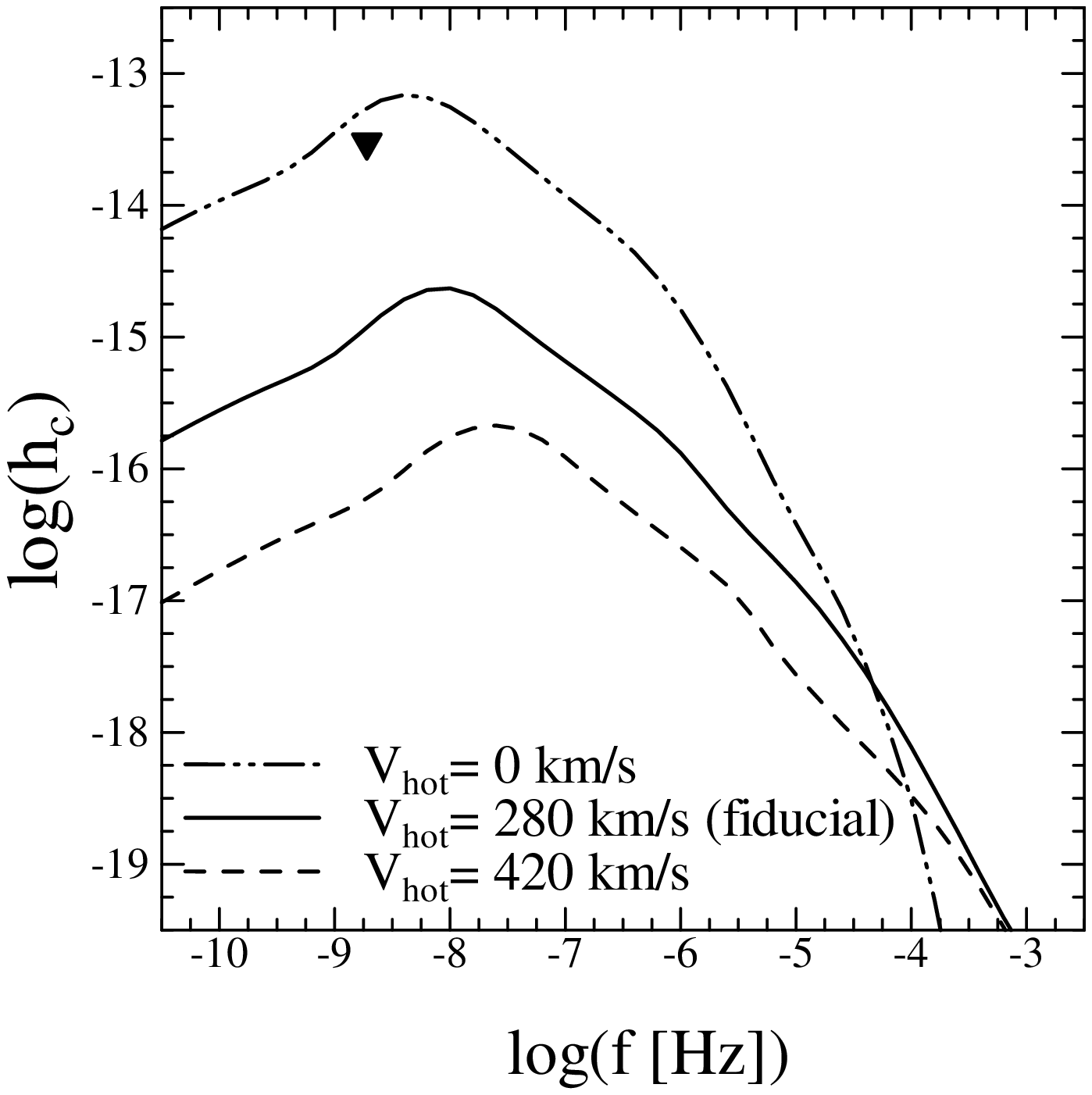}
\caption{Effects of galaxy formation processes on the power spectrum of GWBR from
 SMBH binaries with $e_0=0.8$ and  $f_{p,0}/f_{p, {\rm max}} =
 10^{-3}$. Left panel: For three star formation time scales, $\tau_{*,0}=
 0.5, 1.5$ and $4.5~{\rm Gyr}$. Here $\tau_{*,0}= 1.5~{\rm Gyr}$ is the
 fiducial value. 
Right panel: For three supernovae feedback strengths, $V_{\rm hot}=
 0, 280$ and $420~{\rm km~s}^{-1}$. Here, $V_{\rm hot}= 280~{\rm km~s}^{-1}$
 is the fiducial value. The case  $V_{\rm hot}= 0 ~{\rm km~s}^{-1}$ 
corresponds to no supernova feedback.
In each panel, we also plot the current limit obtained from pulsar timing measurements
 (solid triangle).\cite{lommen02} }
\label{fig07}
\end{figure}

The left panel of Fig. \ref{fig07} displays power spectra of GWBR from SMBH binaries
 for  various star formation time scales, $\tau_{*}^{0}$. Other
 parameters are the same in each case. It is seen that
 for large $\tau_{*}^{0}$, the SFR is small and thus a large amount of 
 cold gas remains in each galaxy. Thus, the masses of
SMBHs become large. Therefore, the amplitude of the power spectrum of
 the GWBR becomes
 large. Moreover, the slope changing frequency moves forward lower frequency,
 because the upper limit frequency, $f_{p, {\rm max}}$, becomes small.  
The right panel of Fig. \ref{fig07} plots power spectra of GWBR from SMBH
 binaries for various supernovae feedback strengths, $V_{\rm hot}$. Other
 parameters are the same in each case.
In the case of no supernovae feedback ($V_{\rm hot} = 0 ~{\rm km~s}^{-1}$), the cold gas is not
heated to a hot gas. Thus, a large amount of cold gas remains in each galaxy, and the
masses of SMBHs become large. By contrast, for the large $V_{\rm hot}$ case, 
the cold gas is greatly heated, and thus the masses of SMBH become small. 
Both panels of Fig. \ref{fig07} indicate that although  the shape of the
 power spectrum of the GWBR from SMBH binaries is affected by the eccentricity of binaries
 for low frequencies, $f \lesssim 1~{\rm nHz}$,
 the shape of the power spectrum for higher frequencies and the overall
amplitude depend strongly on the processes of galaxy formation.

\begin{figure}
\centerline{\includegraphics[width=0.5\textwidth]{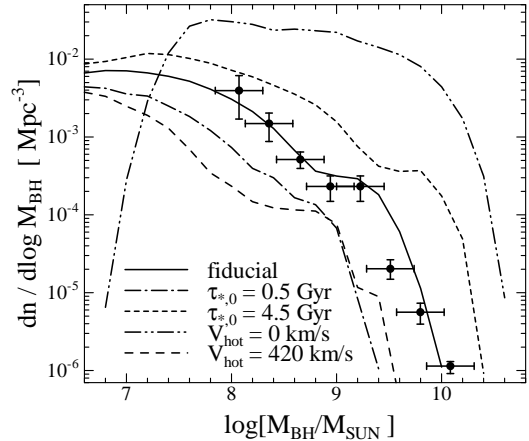}}
\caption{SMBH mass functions of several models at $z=0$. The solid curve represents the
 results of the model with fiducial values, $\tau_{*,0}= 1.5~{\rm Gyr}$ and
 $V_{\rm hot}= 280~{\rm km~s}^{-1}$. 
The dots with error bars present the data for
 the present mass function obtained by Salucci et al.\cite{salucci99}}
\label{fig08}
\end{figure}

We note that the purpose of this subsection is to demonstrate the importance of galaxy formation
processes on the growth of SMBHs and the GWBR from SMBH binaries. Therefore, 
in this subsection, we present cases with {\it extreme parameter values}.
  Results for models with these parameter values are greatly
  inconsistent with observational results, such as galaxy luminosity functions and
the present SMBH mass function. In Fig. \ref{fig08}, we plot
SMBH mass functions at $z=0$. It is seen that only the model with fiducial values is
consistent with the observational results obtained by Salucci et al\cite{salucci99}.

\section{Summary and conclusions}\label{sec.summary}

In this study, we have investigated how orbital eccentricities of binaries
affect power spectra of GWBR from coalescing SMBH binaries.
A compact binary in an eccentric orbit radiates GWs at all integer harmonics of
its orbital frequency. 
Owing to this harmonic radiation, the SED, the power and the timescale
of the GW emission 
 of a binary in an eccentric orbit are 
different from those of a binary in a circular orbit.
Therefore, first, we formulated the power spectrum of GWBR from
cosmological compact binaries in eccentric orbits. Then using this
formulation  and our SA model for galaxy and SMBH formation, we 
calculated the power spectra of GWBR from 
coalescing SMBH binaries in eccentric orbits.

We found that the calculated power spectra of the GWBR from SMBH binaries in
eccentric orbits are
suppressed owing to the harmonic radiation for lower frequencies 
($f \lesssim 1~{\rm nHz}$) if the initial
eccentricity satisfies $e_0 > 0.2$ at $a = 300~R_{\rm S}$.   
The degree of this suppression depends strongly on the initial eccentricity.
 In this paper, for simplicity and in order to clarify the effect of the
 eccentricity, we assumed that all binaries have the
 same initial eccentricity. The results of our study suggest that elucidating
 the initial eccentricity 
distribution is essential for the investigation of the power spectrum
of GWBR from compact binaries. 

Because the  number density of coalescing SMBH
binaries is determined by the processes of galaxy formation, the power spectrum of
the GWBR from SMBH binaries depends strongly on these processes, especially
processes related to star formation, which regulate the amount of cold gas
accreted onto SMBHs. The prediction of 
our model shows that the overall shape and amplitude of
the power spectrum of the GWBR from coalescing
SMBH binaries depend on galaxy formation processes. However,
for low frequencies ($f \lesssim 1~{\rm nHz}$), the shape of the power
spectrum of the GWBR also depends on the initial
eccentricity. Because the pulsar timing measurements can provide limits on
the level of GWBR with frequencies $f \sim 1~{\rm n}- 100~{\rm nHz}$, 
pulsar timing observations, such as PPTA
project, should provide constraints not only on the number density of coalescing SMBH
binaries  but also on the effect of orbital eccentricity on the GWBR from SMBH binaries.

\section*{Acknowledgements}
We thank Dr. K. T. Inoue for useful suggestions.
The numerical computations in this work were partly carried out at the
Astronomy Data Center of the National Astronomical
Observatory of Japan.
This work was supported in part by a Nagasaki University President's
Fund Grant.

\appendix
%\section{First Appendix} %Empty argument \section{} yields `Appendix'. 
%\section{}
%\section{Second Appendix}

\end{document}